\documentclass[11pt]{article}
\usepackage{amsmath, amsthm, amsfonts}
\usepackage[margin=1in,footskip=0.25in]{geometry}
\usepackage{makecell}
\usepackage{multicol}
\usepackage{graphicx}
\theoremstyle{theorem}

\usepackage{hyperref}
\usepackage{cite}

\newtheoremstyle{defi}
  {10pt}          
  {10pt}  
  {\rm}  
  {\parindent}     
  {\bf}  
  {. }    
  { }    
  {}     
\theoremstyle{defi}

\begin{document}

\date{}

\title{\bf Nonlinear electrodynamics, singularities and accelerated expansion}
\author{Gauranga C. Samanta$^{1}$\footnote{gauranga81@gmail.com},
Shantanu K. Biswal$^{2}$, Kazuharu Bamba$^{3}$ \\
 $^{1}$Department of Mathematics,
BITS Pilani K K Birla Goa Campus,
Goa-403726, India,\\
$^{2}$Gandhi Institute for Technological Advancement, Bhubaneswar, India\\
$^{3}$Division of Human Support System, Faculty of Symbiotic Systems Science,\\
Fukushima University, Fukushima 960-1296, Japan \\
}

\maketitle

\begin{abstract}
Non-linear electrodynamics coupled to general relativity is investigated. In general relativity, it is observed that the expansion of the universe is accelerating if the source of the gravitational field is the non-linear electromagnetic field. Moreover, a pure magnetic universe is explored. As a result, it is demonstrated that the magnetic field drives the cosmic expansion to be accelerated. Furthermore, the stability analysis of the present model is developed. In addition, the energy conditions and future singularities are studied in detail.
\end{abstract}



\section{Introduction}

The cosmic microwave background (CMB) and type Ia supernovae observations suggests that, our universe is expanding with acceleration. We may explain this
cosmic acceleration of the universe by introducing a cosmological constant $(\Lambda)$ in the Einstein's Hilbert action with the equation of state
$p=\omega\rho$, where $\omega=-1$ (with $\rho$ and $p$ the energy density and pressure of the universe, respectively). The field with such type of property is called dark energy. To account for this cosmic acceleration, the cosmological constant $(\Lambda)$ is one of the simple and natural candidate, however it faces vital problems of fine-tuning and a large mismatch between theory and observations. We may explain this dark energy in another way by introducing scalar field with proper potential. There has been significant attempts to construct the dark energy models by modifying the Einstein's Hilbert action. This approach is called modified gravity. The several modification of general relativity has been done by the replacement of the Ricci scalar ($R$) in Einstein's Hilbert action by some function of $f(R)$, $f(R, T)$ etc.\cite{Starobinsky, Capozziello, Capozziello1, Carroll, Nojiri, Chiba, Dolgov, Soussa, Olmo, Faraoni, Bamba, Bamba1, Bamba3, Harko, Houndjo, Jamil, Myrzakulov, Sharma, Shabani, Moraes, Ram, Samanta, Samanta1, Samanta2} \bibliographystyle{IEEEtran}
(for reviews on theories of modified gravity as well as
the issue of dark energy, which are studied to explain the late-time cosmic acceleration, see, for example,~\cite{Nojiri:2006ri, Nojiri:2010wj, Book-Capozziello-Faraoni, Capozziello:2011et, Bamba:2012cp, Joyce:2014kja, Koyama:2015vza, Bamba:2015uma, Cai:2015emx, Nojiri:2017ncd}).

All researchers are explaining the expansion of the universe in several ways. At the same time non-linear electrodynamics may solve the problem of early time inflation and late time cosmic acceleration of the universe by without modification of the general relativity. The electromagnetic fields are very strong at the earlier time of the evolution of the universe. If the radiation-dominated stage in the early universe is governed by the Maxwell's equations, then there will be a space like an initial singularity in the past. However, the initial singularity may avoid by modifying the Maxwell's equations in the early stage of the universe.
The finite-time future singularities in various modified gravity theories
have been investigated in detail~\cite{Bamba:2008ut, Bamba:2008hq, Bamba:2009uf, Bamba:2012vg}.
Recently, in Refs.~\cite{Novello, Novello1, Novello2, Lorenci}, it has been shown that the initial singularity can be avoided and have a period of late time cosmic acceleration when the universe field with magnetic field with Lagrangian $L=-\frac{1}{4}F^2+\alpha F^4-\frac{\gamma^8}{F^2}$, where $F^2=F^{\mu\nu}F_{\mu\nu}$ and $\alpha$ and $\gamma$ are constants. At the early stage of the universe, electromagnetic and gravitational fields are very strong, and therefore, the non-linear electromagnetic source may be taken into consideration. The non-linear electrodynamics is a approximation of the Maxwell's theory for weak field and the space-time with non-linear electrodynamics can be applied for strong fields. Hence, the non-linear electrodynamics has been used to make inflation in the early universe~\cite{Salcedo, Camara} and it can also have cosmological contributions as a source of the late-time acceleration of the universe~\cite{Elizalde, Novello4}. The gravity coupling with non-linear electrodynamics may produce negative pressure and that is the cause of acceleration \cite{Novello5, Vollick, Montiel}. Hence there is no need of dark energy components to explain the late time cosmic accelerated expansion. The isotropic and homogeneous cosmological models coupled with electromagnetic Born-Infeld (BI) field is tested with the standard probes of SNIa, GRBs and direct Hubble parameter \cite{Breton}. The cosmological consequences in the existence of the non-minimal coupling between electromagnetic fields and gravity have been explored~\cite{Bamba:2008ja, Bamba:2008xa}. In addition, the influence of the existence of strong magnetic fields on the propagation of gravitational waves~\cite{Bamba:2018cup} and the relationship between the cosmic accelerated expansion and the cosmic magnetic fields~\cite{MRTB} have been studied.

In this paper, we study a model of the non-linear electrodynamics~\cite{Kruglov} coupled with gravitational fields. It is seen that in general relativity,
if the source of the gravitational field is the non-linear electromagnetic field, the accelerated expansion of the universe can be realized. We also
investigate a pure magnetic universe and show that the accelerated expansion
of the universe is driven by the magnetic fields.
Moreover, we analyze the stability of the present model explicitly.
Furthermore, we explore the energy conditions and future singularities in detail.

The organization of the paper is as follows.
In Sec.~II, we explain models of non-linear electrodynamics.
In Sec.~III, we analyze the cosmological solutions in non-linear
electrodynamics. Furthermore, for non-linear electrodynamics,
we examine the energy conditions and future singularities.
Conclusions are finally provided in Sec.~IV.
Throughout the paper, we use the units $c=\hbar=\varepsilon_0=\mu_0=1$.

\section{Non-linear electrodynamics models}
Let us consider the Lagrangian density of nonlinear electrodynamics \cite{Kruglov} is defined as
\begin{equation}\label{1}
  \mathcal{L}_{em}=-\frac{\mathcal{F}}{2\beta\mathcal{F}+1},
\end{equation}
where $\mathcal{F}=\frac{1}{4}F_{\mu\nu}F^{\mu\nu}=\frac{B^2-E^2}{2}$, $F_{\mu\nu}$ is the field strength tensor and $\beta\mathcal{F}$ is a dimensionless quantity. The denominator of the Lagrangian \eqref{1} will not vanish because the strength of the electric field can not reach the value
$E_{max}=\frac{1}{\sqrt{\beta}}$. The energy momentum tensor in \cite{Kruglov} is defined as
\begin{equation}\label{2}
  T_{\mu\nu}=-\frac{1}{(2\beta\mathcal{F}+1)^2}[F^{\alpha}_{\mu}F_{\nu\alpha}-g_{\mu\nu}\mathcal{F}(2\beta \mathcal{F}+1)]
\end{equation}
and it has a nonzero trace. Based on equation \eqref{1}, the scale variance in the nonlinear electrodynamics model is broken and that support for
the negative pressure. We can make the average of the electromagnetic fields which are sources in general relativity \cite{Tolman} to have the isotropy of the Friedman-Robertson-Walker (FRW) space-time.
Hence, here, we use the average values of the electromagnetic fields as follows:
\begin{equation}\label{3}
  <E>=<B>=0, <E_{i}, B_j>=0, <E_i, E_j>=\frac{1}{3}E^2g_{ij}, <B_i, B_j>=\frac{1}{3}B^2g_{ij} \nonumber
\end{equation}

Let us consider the Friedman Robertson Walker $(FRW)$ space-time
\begin{equation}\label{3}
  ds^2=-dt^2+a^2(t)\bigg[\frac{dr^2}{1-kr^2}+r^2d\theta^2+r^2\sin^2\theta d\phi^2\bigg],
\end{equation}
where $k$ is curvature, there are three different values for $k$, such as $-1, 0, 1$. The universe is spatially open for $k=-1$, the universe is spatially closed for $k=1$ and the universe is spatially flat for $k=0$. Where the co-ordinate systems $(r, \theta, \phi)$ are co-moving co-ordinates, i. e. an observer at rest in these coordinates remains at rest.
The Einstein's Hilbert action of general relativity coupled with the nonlinear electromagnetic field described
by the Lagrangian density \eqref{1} is
\begin{equation}\label{4}
  S=\int\frac{\sqrt{-g}R}{2\kappa^2}d^4x+\int\sqrt{-g}\mathcal{L}_{em}d^4x,
\end{equation}
where $R$ is the Ricci scalar and $\kappa^{-1}=M_{pl}$, $M_{pl}\equiv (8\pi G)^{\frac{-1}{2}}$ is the reduced Planck mass, where $G$ is Newton's constant.
Max Planck introduced his famous units of mass, length and time a hundred years ago and constructed exclusively out of the three fundamental constants,
$\hbar=\frac{h}{2\pi}$, $c$ and $G$ \cite{Planckm}, where $\hbar$ is the Planck constant introduced by him only in 1900, $c$ is the velocity of light (leading laws of relativity) and $G$ is the Newtonian gravitational constant. The Planck mass arises very frequently in astrophysics, cosmology, quantum gravity, string theory, etc. The Planck mass is defined as $M_{pl}=\left(\frac{\hbar c}{G}\right)^{\frac{1}{2}}=2.2\times 10^{-5}gm$, however in this paper we have considered $M_{pl}\equiv \left( \frac{1}{8\pi G}\right)^{\frac{1}{2}}$. For gravity, the gravitational constant $G$ is inversely proportional to $M_{pl}^2$, i. e. $G\propto \frac{1}{M_{pl}^2}$. Under weak gravity Plank mass becomes very large. At the Plank energy, all quantum gravitational process
becomes very strong. Further, if G changes with time or with energy, or the gravity coupling scales logarithmically with energy, then we can no longer define the Planck units with constant values at high energies. So the parameter $M_{pl}=\left(\frac{\hbar c}{G}\right)^{\frac{1}{2}}=2.2\times 10^{-5}gm$
becomes energy dependant.
The field equations follow from equation \eqref{4} as
\begin{equation}\label{5}
  R_{\mu\nu}-\frac{1}{2}Rg_{\mu\nu}=-\kappa^2T_{\mu\nu},
\end{equation}
\begin{equation}\label{6}
  \partial_{\mu}\bigg(\frac{\sqrt{-g}F^{\mu\nu}}{(2\beta\mathcal{F}+1)^2}\bigg)=0.
\end{equation}
Where $T_{\mu\nu}=(p+\rho)u_{\mu}u_{\nu}+pg_{\mu\nu}$ and the four velocity vector $u^{\mu}$ is defined as $u^{\mu}=(1, 0, 0, 0)$.
The energy density $(\rho)$ and the pressure $(p)$ are obtained from \eqref{2} as follows:
\begin{equation}\label{7}
  \rho=\frac{E^2}{(2\beta\mathcal{F}+1)^2}+\frac{\mathcal{F}}{2\beta\mathcal{F}+1}
\end{equation}
\begin{equation}\label{8}
  p=\frac{2B^2-E^2}{3(2\beta\mathcal{F}+1)^2}-\frac{\mathcal{F}}{2\beta\mathcal{F}+1}
\end{equation}
\par
The explicit form of the field equations \eqref{5} with the help of space time \eqref{3} having zero curvature (i. e. $k=0$) as follows:
\begin{equation}\label{9}
  3\frac{\dot{a}^2}{a^2}=\kappa^2 \rho
\end{equation}
\begin{equation}\label{10}
  2\frac{\ddot{a}}{a}+\frac{\dot{a}^2}{a^2}=-\kappa^2 p
\end{equation}
By performing equation \eqref{9} and \eqref{10}, we get
\begin{equation}\label{11}
  \frac{\ddot{a}}{a}=-\frac{1}{6}\kappa^2(\rho+3p)
\end{equation}
which is sometimes called the Raychaudhuri equation.
We know, acceleration and deceleration of the universe depends on the sign of $\ddot{a}$. Hence, from equation \eqref{11} we can say that, the universe
will accelerate for $\rho+3p<0$ and decelerate for $\rho+3p>0$. If the linear equation of state between $p$ and $\rho$ is $p=\omega\rho$ holds, then
the $\omega>\frac{-1}{3}$  for deceleration and $\omega<\frac{-1}{3}$ for acceleration.
Let us suppose that the field of nonlinear electrodynamics is the main source of gravity. Here we considered, only magnetic field is important in cosmology, so we can have $E=0$. Thus, the electric field is screened because of the charged primordial plasma, but the magnetic field is not screened \cite{Lemoine}.
According to cosmological principle, $<B_i>=0$. Hence the magnetic field does not induce the directional effects. Performing equations
\eqref{7} and \eqref{8} with $E^2=0$, we get
\begin{equation}\label{12}
  \rho=\frac{B^2}{2(\beta B^2+1)}
\end{equation}
\begin{equation}\label{13}
  \rho+p=\frac{2B^2}{3(\beta B^2+1)^2}
\end{equation}
\begin{equation}\label{14}
  \rho+3p=\frac{B^2(1-\beta B^2)}{(\beta B^2+1)^2}
\end{equation}
The equation of continuity is obtained as
\begin{equation}\label{15}
  \dot{\rho}+3H(p+\rho)=0
\end{equation}
For accelerating universe, we need $\rho+3p<0$. Hence, from equation \eqref{14} we observe that $1-\beta B^2<0$, i. e. $1<\beta B^2$ is required to explain cosmic acceleration of the universe. This inequality will happen at epoch under strong magnetic fields.
Therefore, the inequality $\rho+3p<0$ can be satisfied and drives the accelerated expansion of the universe under nonlinear electrodynamics and the magnetic field scenario.

\section{Cosmological solutions}
In this section, we would like to investigate the dynamics of the energy density $(\rho)$, pressure $(p)$, electromagnetic field $(B)$, scale factor $(a)$
and deceleration parameter $(q)$ of the model.

From the equations \eqref{9}, \eqref{10}, \eqref{12} and \eqref{13}, we can have
\begin{equation}\label{16}
  B^2=\frac{1}{\beta}\frac{1-q}{1+q},
\end{equation}
where $q$ is deceleration parameter and defined as follows
\begin{equation}\label{17}
  q=-\frac{a\ddot{a}}{\dot{a}^2}
\end{equation}
We can certainly see that, right side of the equation \eqref{16} must be positive. Hence $q$ varies from -1 to 1, i. e. $-1<q<1$. There is a singularity in
equation \eqref{16} of electromagnetic field for $q=-1$, so $q\ne -1$. Hence, we confirmed from the above equation \eqref{16} that our universe does not follow de Sitter expansion. For $q=1$, we get $B=0$, i. e. our universe does not contain any electromagnetic field. Since observational data suggests that, the positive deceleration parameter indicates the deceleration phase of the universe. Hence we may conclude that without electromagnetic field, the only general theory of relativity is not a capable candidate to explain the current accelerated expansion of the universe, so $q\ne 1$ is acceptable. Therefore the electromagnetic field reveals that the range of the deceleration parameter is $-1<q<0$, which is absolutely compatible with our current observations.

Dividing equations \eqref{10} by \eqref{9}, we can get $-\frac{p}{\rho}=\frac{1}{3}+\frac{2}{3}\frac{a\ddot{a}}{\dot{a}^2}$. Since $q=-\frac{a\ddot{a}}{\dot{a}^2}$, so we can write $\frac{p}{\rho}=-\frac{1}{3}+\frac{2}{3}q$. The usual assumption in cosmology is that there is a
unique pressure associated with each density, so that $p\propto p(\rho)$. Such a relation is known as the equation of state. The simplest one is
$p=\omega \rho$, where $\omega$ is constant and this $\omega$ is called the equation of state parameter. However, in this paper we have considered
$\omega$ is time dependent rather than constant i. e. $p=\omega (t)\rho$. Hence, we can write the deceleration parameter in terms of the time dependent
equation of state parameter as follows

\begin{equation}\label{18}
  q=\frac{1+3\omega(t)}{2}.
\end{equation}

Let us introduce a non-linear function $\eta=\ln a$. Now, we can write $\frac{\ddot{\eta}}{\dot{\eta}^2}=\frac{-3}{2}(\omega(t)+1)=-(q(t)+1)$, which
suggests to define a function $f(t)$, and such $f(t)$ can be defined as $f(t)=\frac{3}{2}(\omega(t)+1)=-q(t)-1$. Now, we can write
$\omega(t)=-1+\frac{2}{3}f(t)$ and $q(t)=-1-f(t)$. This transformations help us to reduce the order of the differentia equation. Hence, we can write
$f(t)=-\frac{\ddot{\eta}}{\dot{\eta}^2}= \left(\frac{1}{\dot{\eta}}\right)^{.}$, which implies $\dot{\eta}=\left(\int f(t) dt+c_1\right)^{-1}$. Now, we can write scale factor, energy density and pressure in quadrature form.
\begin{equation}\label{19}
  a(t)=a_0\exp\left(\int_{t_0}^{t}\left(\int f(t)dt+c_1\right)^{-1}dt\right),
\end{equation}
\begin{equation}\label{20}
  \rho=\frac{3}{\kappa^2}\left(\int_{t_0}^{t}f(t)dt+c_1\right)^{-2}
\end{equation}
   and
   \begin{equation}\label{21}
     p=\frac{1}{\kappa^2}\frac{3(f(t)-1)}{\left(\int_{t_0}^{t}f(t)dt+c_1\right)^2}.
   \end{equation}

   From the equations \eqref{13} and \eqref{15}, one can have the relation between the electromagnetic field $(B(t))$ and the scale factor $(a(t))$ as
   \begin{equation}\label{22}
     B\propto \frac{1}{a^2}.
   \end{equation}
    This shows that, the evolution of the magnetic field follows inverse square law of the scale factor $(a)$.
   The dynamics of the model depends on the behavior of $f(t)$. Let us assume that $f(t)$ follows Puiseux series expansion of time around $t=0$.
   \begin{center}
     $f(t)=f_0t^{\gamma_0}+f_1t^{\gamma_1}+f_2t^{\gamma_2}+\cdots  + f_nt^{\gamma_n}+\cdots , ~~~~\gamma_0<\gamma_1<\cdots$.
   \end{center}
Now, we try to get the explicit form of the scale factor $(a)$, the energy density $(\rho)$ and the pressure $(p)$ at lowest order in cosmic time $t$,
\begin{equation}\label{23}
  \eta (t)=\begin{cases}
             -\frac{\gamma_0+1}{\gamma_0f_0}t^{-\gamma_0}+\cdots, & \mbox{if } \gamma_0\ne -1, 0 \\
             -\frac{t}{f_0}-\frac{f_0+f_1}{2f_0^2}t^2+\cdots, & \mbox{if } \gamma_0=-1, |t|\le 2 \\
             \frac{\ln t}{f_0}-\frac{f_1}{2g_0^2}t+\cdots, & \mbox{if } \gamma_0=0
           \end{cases}
\end{equation}
\begin{equation}\label{24}
  a(t)=\begin{cases}
             \exp\left(-\frac{\gamma_0+1}{\gamma_0f_0}t^{-\gamma_0}+\cdots \right), & \mbox{if } \gamma_0\ne -1, 0 \\
             \exp\left(-\frac{t}{f_0}-\frac{f_0+f_1}{2f_0^2}t^2+\cdots \right), & \mbox{if } \gamma_0=-1, |t|\le 2 \\
             \exp\left(\frac{\ln t}{f_0}-\frac{f_1}{2g_0^2}t+\cdots \right), & \mbox{if } \gamma_0=0
           \end{cases}
\end{equation}

\begin{equation}\label{25}
  \rho(t)=\begin{cases}
            3\left(\frac{\gamma_0+1}{f_0}\right)^2t^{-2(\gamma_0+1)}+\cdots, & \mbox{if } \gamma_0\ne -1, 0 \\
            \frac{3}{f_0^2 (\ln t)^2}+\cdots, & \mbox{if } \gamma_0=-1 \\
            \frac{3t^{-2}}{f_0^2}+\cdots, & \mbox{if } \gamma_0=0
          \end{cases}
\end{equation}
\begin{equation}\label{26}
  p(t)=\begin{cases}
         \frac{3}{f_0 t (\ln t)^2}+\cdots , & \mbox{if } \gamma_0=-1 \\
         \frac{3(f_0-1)}{f_0^2}t^{-2}+\cdots, & \mbox{if } \gamma_0=0 \\
         \frac{3(\gamma_0+1)^2}{f_0}t^{-(\gamma_0+2)}+\cdots , & \mbox{if } \gamma_0\ne -1, \gamma_0<0 \\
         -3\left(\frac{\gamma_0+1}{f_0}\right)^2t^{-2(\gamma_0+1)}+\cdots, & \mbox{if } \gamma_0>0.
       \end{cases}
\end{equation}
The following observations are made from the above expressions of the scale factor, energy density and pressure of the nonlinear electromagnetic model.

\subsection{Stability of the model}
The particles of the universe has been classified into three classes, mainly, sub-luminal, luminal and supper-luminal. The particles move with slow speed compare to the speed of light are called sub-luminal particles, for example electrons and neutrons. The particles move with exactly same as the speed of light are called luminal particles, for example photon and graviton. The particles move with faster than the speed of light are called super-luminal particles or tachyons. There are two possibilities for the existence of supper-luminal particles: either they do not exist or, if they do, then they do not interact with an ordinary matter. If the speed of the sound is less than the local light speed, $c_s\le 1$, then only we can say causality may not be violated. The positive square sound speed $(c_s^2>0)$ is necessary for the classical stability of the universe. The speed of sound is defined as $\frac{dp}{d\rho}=c_s^2$ \cite{Ellis}. From the equations \eqref{7} and \eqref{8}, we obtain the speed of sound (with $E=0$)
\begin{equation}\label{27}
  \frac{dp}{d\rho}=c_s^2=-\frac{11\beta B^2+3}{3(\beta B^2+1)}
\end{equation}
From the equation \eqref{27}, one can verify that our universe follows the classical stability condition for the scale factor  $a>\left(\frac{14\beta}{6}\right)^{\frac{1}{4}}\sqrt{B_0}$, however the universe follows instability condition i. e. the universe may contains some abnormal (something not normal) matters (may be tachyons like matters) if the scale factor $a<\left(\frac{14\beta}{6}\right)^{\frac{1}{4}}\sqrt{B_0}$. Hence we may assume that the universe was filled up with some abnormal matters or tachyons in early epoch i. e. from the big bang to inflationary stage and as a result it causes inflation.

The following observations are made for the sound speed and causality of the model based on the evolution of the energy density and pressure given in equations \eqref{25} and \eqref{26}.

\begin{itemize}
  \item For $\gamma_0=-1$, the sound speed is obtained as $c_s^2=\frac{dp}{d\rho}=\frac{\ln t}{2t}+\frac{1}{t}$. The model satisfies the stability conditions for $e^{-1}<t<e^{(t-1)}$ and $t>1$. The model is stable and therefore causality cannot be violated during this period. For the existence of abnormal matters the $c_s^2$ must be greater than one i. e. $(c_s^2>1)$, which implies that $t>e^{2(t-1)}$, however this is not feasible. Hence we could not get any theoretical evidence for the existence of abnormal matters in the present universe. Moreover, there may be a chance for the existence of the abnormal matters before inflation.
   \item For $\gamma_0=0$, the sound speed is obtained as $c_s^2=\frac{dp}{d\rho}=f_0-1$. The model satisfies stability condition for $1<f_0<2$. For $f_0=2$, indicates the existence of ordinary matters and for $f_0>2$, indicates the existence of tachyons in the universe respectively.

   \item For $\gamma_0 \ne 0, -1$ and $\gamma_0<0$, the sound speed is obtained as $c_s^2=\frac{dp}{d\rho}=\frac{(\gamma_0+2)t^{\gamma_0}}{2(\gamma_0+1)}$. $t<\left(\frac{2(\gamma_0+1)}{\gamma_0+2}\right)^{\frac{1}{\gamma_0}}$ indicates for the existence of ordinary matters (i.e. $c_s^2<1$), $t>\left(\frac{2(\gamma_0+1)}{\gamma_0+2}\right)^{\frac{1}{\gamma_0}}$ indicates for the existence of abnormal matters (i.e. $c_s^2>1$), $t=\left(\frac{2(\gamma_0+1)}{\gamma_0+2}\right)^{\frac{1}{\gamma_0}}$ indicates for the existence of ordinary matters  (i.e. $c_s^2=1$) in the universe respectively. The condition $c_s^2>0$ is satisfied for
       $\gamma_0 \in (-1, 0)\cup (-\infty, -2)$. The model maintains causality for very small time period i. e.
       $t\in\bigg[0, \left(\frac{2(\gamma_0+1)}{\gamma_0+2}\right)^{\frac{1}{\gamma_0}}\bigg]$, however the model violate causality for very long time period i. e. $t\in \bigg[\left(\frac{2(\gamma_0+1)}{\gamma_0+2}\right)^{\frac{1}{\gamma_0}}, \infty\bigg]$. Hence in this case either the model is not physically realistic or indicates the existence of abnormal matters in present and future universe. The presence of abnormal matters of the universe causes the accelerated expansion of the universe.

   \item  For $\gamma_0 \ne 0, -1$ and $\gamma_0>0$, the sound speed is obtained as $c_s^2=\frac{dp}{d\rho}=-1$, which is either not acceptable or indicates the existence of non-normal matter in the universe.

\end{itemize}

\subsection{Energy conditions}

It is sensible to hope that the stress-energy tensor should satisfy certain conditions, such as positivity of the energy density and dominance of the energy density over the pressure. In general relativity the energy conditions are divided into four parts \cite{Hawking} such as:
\begin{itemize}
  \item [I] Weak Energy Condition (WEC).
  \item [II] Null Energy Condition (NEC).
  \item [III] Strong Energy Condition (SEC).
  \item [IV] Dominant Energy Condition (DEC).
\end{itemize}
The following observations are made, based on the evolution of the energy density and pressure from the equations \eqref{25} and \eqref{26}.
\begin{itemize}
  \item \textbf{Weak Energy Condition:} The weak energy condition states that the energy density for any kind of matter distribution is non-negative as measured by any observer in space-time. The energy momentum tensor measured by an observer at each $p\in \mathbb{M}$ (where $p$ is the point and $\mathbb{M}$ is the four dimensional manifold) with any time-like vector $u^{\mu}$ is $T_{\mu\nu}u^{\mu}u^{\nu}\ge 0$. By assuming isotropic pressure, which implies $\rho\ge 0, ~~ \rho+p\ge 0$. From the equations \eqref{25} and \eqref{26} we can see that at any point $\rho\ge 0$ for all observer. Similarly at any point $p+\rho\ge 0$ for all observer. Hence, the model satisfies weak energy condition.
  \item \textbf{Null Energy Condition:} The statement of null energy condition is same as the weak form, except that $u^{\mu}$ is replaced by an arbitrary, future-directed null vector $k^{\mu}$. Hence $T_{\mu\nu}k^{\mu}k^{\nu}\ge 0$ is follow the null energy condition inequality. By assuming isotropic pressure, the $NEC$ implies $p+\rho\ge 0$. Notice that the $WEC$ implies the $NEC$. Therefore, the model also satisfies $NEC$.
  \item \textbf{Strong Energy Condition:} The $SEC$ is defined as $\left(T_{\mu\nu}-\frac{1}{2}Tg_{\mu\nu}\right)u^{\mu}u^{\nu}$, or $T_{\mu\nu}u^{\mu}u^{\nu}\ge -\frac{1}{2}T$, where $u^{\mu}$ is any future-directed, normalized, time-like vector. By assuming isotropic pressure, the explicit form of the $SEC$ is as follows $\rho+3p\ge 0, ~~\rho+p\ge 0$. It is noted that the $SEC$ does not imply the $WEC$ and $NEC$. From equations \eqref{25} and \eqref{26}, we can have
      \begin{equation}\label{28}
        \rho+3p=-6\left(\frac{\gamma_0+1}{f_0}\right)^2t^{-2(\gamma_0+1)}, ~\mbox{ for} \gamma_0>0
      \end{equation}
      From equation \eqref{28}, it is observed that the $SEC$ does not satisfy for $\gamma_0>0$. The violation of $SEC$ indicates that the universe contains some non-normal matter (may be supper-luminal particles, for example tachyons). Also, we observed that the model does not satisfies the sound speed and causality conditions.
  \item \textbf{Dominant Energy Condition:} The $DEC$ states that matter should flow along time-like or null world lines. Precisely, if $u^{\mu}$ is an arbitrary, future directed, time-like vector field, then $-T_{\mu}^{\nu}u^{\mu}$ is a future directed, time-like or null, vector field. The matter's momentum density measured by any observer is $-T_{\mu}^{\nu}u^{\mu}$ with four velocity vector $u^{\mu}$, and this is required to be time-like or null. The explicit form of the $DEC$ is $\rho\ge 0, ~~ \rho\ge |p|$. From the equations \eqref{25} and \eqref{26}, it is found that
      \begin{equation}\label{29}
        \rho=\frac{3}{f_0^2(\ln t)^2} \mbox{ for} ~~\gamma_0=-1
      \end{equation}
      \begin{equation}\label{30}
        p=\frac{3}{f_0 t(\ln t)^2} \mbox{ for}~~ \gamma_0=-1
      \end{equation}
      From equations \eqref{29} and \eqref{30}, it is observed that $\rho\ge |p|$ for $t>1$, whereas $\rho\le |p|$ for $0<t<1$. So $DEC$ does not satisfy for $t<1$, whereas it satisfies for $t>1$. Violating the $DEC$ is normally allied with either a large negative cosmological constant or super-luminal particles. Violating the $SEC$ but not the $WEC$,$NEC$ and $DEC$ is associated with either positive cosmological constant or inflationary epoch. Violating $DEC$ and $WEC$ is associated with negative cosmological constat.
\end{itemize}

\subsection{Singularities:}
The universe has been made up with large number of different species of matter fields. In reality, it is complicated to describe the exact energy momentum tensor even if one know the precise form of the each matter and equation of motion governing it. In fact, we don't have much idea about the behavior of matter under extreme conditions of density and pressure with time. Hence, one may predict the occurrence of singularities in the universe in general relativity. Therefore, in this section, we classify the finite time singularity in the following way:

\begin{itemize}
  \item [I.] \textbf{Type I Singularity:} If the equation of state is less than $-1$ in the context of general relativity, then the universe reaches a singularity at finite time, along with the null energy condition $p+\rho\ge 0$ is violated and this type of singularity is called the big-rip singularity \cite{Nojiri1}. In type I singularity, the scale factor, the energy density and the absolute pressure of the universe are blows up for some finite time i. e. $a\to \infty$, $\rho\to \infty$, $|p|\to\infty$ as $t \to t_{f} $.
  \item [II.] \textbf{Type II Singularity:} In type-II, only the absolute value of the isotropic pressure is blows up, moreover the scale factor and the energy density of the universe is finite for finite time t, i. e.  $a\to a_f$, $\rho\to \rho_f$ and $|p|\to \infty$ as $t\to t_f$.
  \item [III.] \textbf{Type III Singularity:} In type-III, the energy density and the isotropic pressure of the universe is blows up, however scale factor
  is finite for finite time t, i. e. $\rho\to\infty$, $|p|\to\infty$ and $a\to a_f$ as $t\to t_f$.
  \item [IV] \textbf{Type IV Singularity:} In type-IV, the energy density and isotropic pressure are conversing to zero and scale factor is finite for
  finite time, i. e. $\rho\to 0$, $|p|\to 0$ and $a\to a_f$ as $t\to t_f$.
\end{itemize}
Here $t_f$, $a_f\ne 0$ and $\rho_f$ are constants. Subsequently, several authors discussed different types of future singularities at finite time, however the null energy condition is not violated
\cite{Barrow, Stefancic, Brevik, Dabrowski, Bouhmadi}. Recently, Fernandez-Jambrina and Lazkoz \cite{Lazkoz} studied detail classifications of singularities and future behavior of the universe in FLRW cosmological models by assuming nonlinear equation of state $(p=f(\rho))$.

Based on the evolution of the scale factor, energy density and pressure given in equations \eqref{24}, \eqref{25} and \eqref{26}, we made the following observations:
\begin{itemize}

  \item \textbf{For $\gamma_0>0$:} both $p$ and $\rho$ diverge at $t=0$, i. e. $|p|\to \infty$ and $\rho\to\infty$, because the equations \eqref{25} and \eqref{26} contain $t^{-2(\gamma_0+1)}$ term. The scale factor $a$ diverges, i. e. $a\to\infty$ provided $f_0<0$ and $a\to 0$ provided $f_0>0$ at $t=0$. The equation of state parameter is defined as $\omega=-1+\frac{2}{3}f(t)$, which is tends to $-1$ at $t=0$. So, it indicates that the model contains type-III singularity for $f_0>0$, however the singularity for $f_0<0$ has not been considered before in the previous frameworks.

  \item \textbf{For $\gamma_0=0$:} both $p$ and $\rho$ diverge at $t=0$, i. e. $|p|\to \infty$ and $\rho\to\infty$, because the equations \eqref{25} and \eqref{26}
  contain $t^{-2}$ term. The scale factor $a$ diverges, i. e. $a\to\infty$ provided $f_0<0$ and $a\to 0$ provided $f_0>0$ at $t=0$. We have defined the equation of state parameter in this work as $\omega=-1+\frac{2}{3}f_0$, which is less than $-1$, i. e. $\omega<-1$ for $f_0<0$ and $\omega>-1$ for $f_0>0$. Hence our model contains type-I or Big Rip singularity for $f_0<0$, which occurs at the phantomlike equation of state, i. e. $\omega<-1$. However, for $f_0>0$ the scale factor becomes zero and the universe collapse, this is called Big Crunch singularities. The Big Crunch is one imaginable situation for the ultimate fate of the universe, in which the expansion of space i. e. the scale factor gradually decreases to zero and the universe re-collapses.

  \item \textbf{For $\gamma_0\in (-1, 0)$:} The isotropic pressure $(p)$, the energy density $(\rho)$ and the equation of state parameter $(\omega)$ diverge at $t=0$, however the scale factor $(a)$ vanishes. These are called type-III, Big Freeze of finite scale factor singularities.

  \item \textbf{For $\gamma_0=-1$:} The isotropic pressure $(p)$ and the energy density $(\rho)$ diverge at $t=0$, whereas the scale factor $(a)$ is non zero constant. These are called type-III singularities.

  \item \textbf{For $\gamma_0\in (-2, -1)$:} The energy density $(\rho)$ vanishes, the scale factor $(a)$ is finite, whereas the isotropic pressure $(p)$ diverges to $\infty$ at $t=0$. These are type-II singularities.

  \item \textbf{For $\gamma_0=-2$:} The energy density $(\rho)$ vanishes at $t=0$, the isotropic pressure $(p)$ and the scale factor $(a)$ are finite, whereas the equation of state parameter $(\omega)$ diverges. These are generalized sudden singularities.

 \item \textbf{For $\gamma_0<-2$:} In this case both pressure $(p)$ and energy density $(\rho)$ vanish at $t=0$,
the scale factor $(a)$ becomes constant, whereas the equation of state parameter $(\omega)$ diverges. There are called type-IV or w-type singularities.

\end{itemize}

\section{Conclusions}

The scale of the magnetic fields generated by the introduction of nonlinear electrodynamics is much smaller than the cosmic magnetic fields in galaxies,
whose scale is about 10 kpc. Such strong magnetic fields originating from the nonlinearity of electrodynamics can be cascade along with the cosmic expansion, so that the magnetic strength can be weaker and the scale of the magnetic fields can be larger. Since the origin of the cosmic magnetic fields such as galactic magnetic fields is not yet understood very well (although there are a number of proposals), the nonlinearity of the electrodynamics can be regarded as a possible mechanism to produce the cosmic magnetic fields which are observed in various astronomical objects like galaxies and the clusters of galaxies at the present time.

In the present paper, we have explored the non-linear electrodynamics
with the coupling to gravitational fields. It has been found that
for general relativity, when the source of the gravitational field is
the non-linear electromagnetic field, the cosmic acceleration is
accelerating. In a pure magnetic universe, it has also been shown that
the accelerated expansion of the universe is driven by the magnetic fields.
Furthermore, we have investigated the stability analysis for
the present model. In additioin, we have studied the energy conditions and future singularities in detail.

 From Eq.~\eqref{11}, it is confirmed that $\rho+3p$ must be less than zero to explain the accelerated expansion of the universe. Subsequently, Eq.~\eqref{14} specifies $\beta B^2>1$ is essential to account for the cosmic accelerated expansion. This inequality will take place under strong magnetic fields. Therefore, the inequality $\rho+3p<0$ can be satisfied and drives accelerated expansion of the universe under the framework of nonlinear electrodynamics.

From Eq.~\eqref{16}, it is shown that the deceleration parameter $q$ cannot be $-1$ (i.e. $q\ne -1$) and hence our universe does not follow the de Sitter expansion. However, it is observed that for $q=1$, we get $B=0$, i.e. our universe does not contain any electromagnetic field.
The positive deceleration parameter indicates the deceleration phase of the universe. Therefore, it may be concluded that without electromagnetic field, only general relativity is not a capable candidate to explain the current accelerated expansion of the universe, so $q\ne 1$ is acceptable. Thus, the electromagnetic field reveals that the range of the deceleration parameter is $-1<q<0$, which is absolutely compatible with our current observations.

From Eq.~\eqref{25} and \eqref{26}, it is observed that the equation of state parameter $\omega= \frac{f_0}{t}$ is less than zero for $\gamma_0=-1$ and $f_0<0$. Since $f_0$ is an arbitrary constant, so without loss of generality, we can assume $f_0=\frac{-t_0}{3}$, this implies $\omega=\frac{-t_0}{3t}$. The behavior of the $\omega$ for different range of $`t'$ is given in the table below:
\begin{center}
  \begin{tabular}{|c|c|c|}
    \hline
    $\omega=\frac{p}{\rho}=\frac{-t_0}{3t}$ & Range of `t' & Evolution of the universe \\
    \hline
    $\omega\to-\infty$ & $t\to 0$ & Inflationary era \\
    \hline
    $\omega<-1$ & $0<t<\frac{t_0}{3}$ & Universe expanding in accelerating way \\
    \hline
    $\omega=-1$ & $t=\frac{t_0}{3}$ & Dominated by cosmological constant \\
    \hline
    $-1<\omega<-\frac{1}{3}$ & $\frac{t_0}{3}<t<t_0$ & Phantom era \\
    \hline
    $\omega>\frac{-1}{3}$ & $t>t_0$ & Expanding \\
    \hline
    $\omega\to 0$ & $t\to\infty$ & Dust universe \\
    \hline
  \end{tabular}
\end{center}
From the above table, we can analyze that the EoS parameter $\omega$ tends to negative infinity at the beginning of the universe. This indicates that the universe may occupy with full of abnormal matters and that perhaps causes the initial inflation of the universe. Afterwards, the EoS parameter gradually increases and tends to negative one i. e. $\omega\to -1$ at certain times, during this period $(0<t<\frac{t_0}{3})$ universe experienced an accelerated expansion. During the period $\frac{t_0}{3}<t<t_0$, the EoS parameter lies between $-1<\omega<\frac{-1}{3}$, this indicates the phantom phase of the universe. Subsequently, the EoS parameter tends to zero at infinite time, which indicates that the universe end up with dust.  
  
From Eq.~\eqref{25} and \eqref{26}, we can also find $\omega=\begin{cases}
                                                               f_0-1, & \mbox{if } \gamma_0=0 \\
                                                               f_0t^{\gamma_0}, & \mbox{if } \gamma_0<0 \\
                                                               -1, & \mbox{if } \gamma_0>0.
                                                             \end{cases}$

 Hence, it is observed that for $\gamma_0=0$, the equation of state parameter $\omega$ is less that -1 if $f_0<0$ and $\omega\to -1$, if $f_0\to 0$
 throughout the evolution of the universe. For $\gamma_0<0$, $\omega=f_0t^{\gamma_0}$ which is less than zero if $f_0<0$ and other behavior of $\omega$ is same as the behavior of $\omega$ given in the above table. However, $\omega=-1$ for $\gamma_0>0$ throughout the evolution of the universe, this confirms that the acceleration of the universe is in accelerating way.

As a result, it has been seen that the accelerated expansion of the universe can be explained through the non-linear electrodynamics coupled to gravity.
The stability analysis of the model has also been executed.
It has been revealed that there is no theoretical evidence for the existence of super-luminal particles in the present universe.
Furthermore, it has been verified that there may be a possibility for the existence of the super-luminal particles before inflation.\\

\textbf{Acknowledgement:} The first author G. C. Samanta is extremely thankful to Council of Scientific and Industrial Research (CSIR),
Govt. of India, for providing financial support \textbf{(Ref. No. 25(0260)/17/EMR-II)} for carrying out the research work. Furthermore, the work of KB was partially supported by the JSPS KAKENHI Grant Number JP25800136 and Competitive Research Funds for Fukushima University Faculty (18RI009).


\begin{thebibliography}{}

\bibitem{Starobinsky} A. A. Starobinsky, Phys. Lett. B \textbf{91}, 99 (1980).

\bibitem{Capozziello}S. Capozziello, Int. J. Mod. Phys. D \textbf{11}, 483 (2002).

\bibitem{Capozziello1}S. Capozziello, V. F. Cardone, S. Carloni, and A. Troisi, Int. J. Mod. Phys. D \textbf{12}, 1969 (2003).

\bibitem{Carroll}S. M. Carroll, V. Duvvuri, M. Trodden, and M. S. Turner, Phys. Rev. D \textbf{70}, 043528 (2004).

\bibitem{Nojiri}S. Nojiri and S. D. Odintsov, Phys. Rev. D \textbf{68}, 123512 (2003).

\bibitem{Chiba} T. Chiba, Phys. Lett. B \textbf{575}, 1 (2003).

\bibitem{Dolgov} A. D. Dolgov and M. Kawasaki, Phys. Lett. B \textbf{573}, 1 (2003).

\bibitem{Soussa} M. E. Soussa and R. P. Woodard, Gen. Rel. Grav. \textbf{36}, 855 (2004).

\bibitem{Olmo} G. J. Olmo, Phys. Rev. D \textbf{72}, 083505 (2005).

\bibitem{Faraoni} V. Faraoni, Phys. Rev. D \textbf{74}, 023529 (2006).

\bibitem{Bamba} K. Bamba, S. Nojiri, S. D. Odintsov and D. S. Gomez, Phys. Rev. D \textbf{90}, 124061 (2014).

\bibitem{Bamba1}
  K.~Bamba, S.~Nojiri and S.~D.~Odintsov,
  Phys.\ Lett.\ B {\bf 698}, 451 (2011).


\bibitem{Bamba3} K. Bamba, C. Q. Geng and Chung-Chi Lee, JCAP \textbf{1011}, 001 (2010).

\bibitem{Harko}T. Harko, F. S. N. Lobo, S. Nojiri and S. D. Odintsov, Phys. Rev. D \textbf{84}, 024020 (2011).

\bibitem{Houndjo}M. J. S. Houndjo, Int. J. Mod. Phys. D \textbf{21}, 1250003 (2012).

\bibitem{Jamil} M. Jamil, D. Momeni, M. Raja and R. Myrzakulov, Eur. Phys. J. C \textbf{72}, 1999 (2012).

\bibitem{Myrzakulov}R. Myrzakulov, Eur. Phys. J. C \textbf{72}, 2203 (2012).

\bibitem{Sharma} U. K. Sharma and A. Pradhan, Int. J. Geom. Meth. Mod. Phys. \textbf{15}, 1850014 (2017).

\bibitem{Shabani} H. Shabani and A. H. Ziaie, Int. J. Mod. Phys. A \textbf{33}, 1850050 (2018).

\bibitem{Moraes}P. H. R. S. Moraes, Eur.Phys.J. C \textbf{75}, 168 (2015).

\bibitem{Ram}S. Ram, S. Chandel,  Astrophys. Space Sci. \textbf{355}, 195 (2015).

\bibitem{Samanta} G. C. Samanta, Int. J. Theor. Phys. \textbf{52}, 2647 (2013).

\bibitem{Samanta1} G. C. Samanta, Int. J. Theor. Phys. \textbf{52}, 2303 (2013).
\bibitem{Samanta2} G. C. Samanta and S. N. Dhal, Int. J. Theor. Phys. \textbf{52}, 1334 (2013).

\bibitem{Nojiri:2006ri}
S.~Nojiri and S.~D.~Odintsov,
eConf C {\bf 0602061} (2006) 06
[Int.\ J.\ Geom.\ Meth.\ Mod.\ Phys.\ {\bf 4}, 115 (2007)].

\bibitem{Nojiri:2010wj}
S.~Nojiri and S.~D.~Odintsov,
Phys.\ Rept.\ {\bf 505}, 59 (2011).

\bibitem{Book-Capozziello-Faraoni}
S.~Capozziello and V.~Faraoni,
\textit{Beyond Einstein Gravity}
(Springer, Dordrecht, 2010).

\bibitem{Capozziello:2011et}
S.~Capozziello and M.~De Laurentis,
Phys.\ Rept.\ {\bf 509}, 167 (2011).

\bibitem{Bamba:2012cp}
  K.~Bamba, S.~Capozziello, S.~Nojiri and S.~D.~Odintsov,
  Astrophys.\ Space Sci.\  {\bf 342}, 155 (2012).

\bibitem{Joyce:2014kja}
  A.~Joyce, B.~Jain, J.~Khoury and M.~Trodden,
  Phys.\ Rept.\  {\bf 568}, 1 (2015).

\bibitem{Koyama:2015vza}
  K.~Koyama,
  Rept.\ Prog.\ Phys.\  {\bf 79}, 046902 (2016).

\bibitem{Bamba:2015uma}
  K.~Bamba and S.~D.~Odintsov,
  Symmetry {\bf 7}, 220 (2015).

\bibitem{Cai:2015emx}
  Y.~F.~Cai, S.~Capozziello, M.~De Laurentis and E.~N.~Saridakis,
  Rept.\ Prog.\ Phys.\  {\bf 79}, 106901 (2016).

\bibitem{Nojiri:2017ncd}
  S.~Nojiri, S.~D.~Odintsov and V.~K.~Oikonomou,
  Phys.\ Rept.\  {\bf 692}, 1 (2017).

\bibitem{Bamba:2008ut}
  K.~Bamba, S.~Nojiri and S.~D.~Odintsov,
  JCAP {\bf 0810}, 045 (2008).

\bibitem{Bamba:2008hq}
  K.~Bamba, C.~Q.~Geng, S.~Nojiri and S.~D.~Odintsov,
  Phys.\ Rev.\ D {\bf 79}, 083014 (2009).

\bibitem{Bamba:2009uf}
  K.~Bamba, S.~D.~Odintsov, L.~Sebastiani and S.~Zerbini,
  Eur.\ Phys.\ J.\ C {\bf 67}, 295 (2010).

\bibitem{Bamba:2012vg}
  K.~Bamba, R.~Myrzakulov, S.~Nojiri and S.~D.~Odintsov,
  Phys.\ Rev.\ D {\bf 85}, 104036 (2012).

\bibitem{Novello} M. Novello, S. E. Perez Bergliaffa, Phys. Rept. \textbf{463}, 127 (2008).

\bibitem{Novello1} M. Novello, E. Goulart, J.M. Salim and S.E. Perez Bergliaffa, Class. Quant. Grav. \textbf{24}, 3021 (2007).

\bibitem{Novello2} M. Novello, S. E. Perez Bergliaffa and J. Salim, Phys. Rev. D \textbf{69}, 127301 (2004).

\bibitem{Lorenci} V. A. De Lorenci, R. Klippert, M. Novello, J. M. Salim, Phys. Rev. D \textbf{65}, 063501 (2002).

\bibitem{Salcedo} R. Garca-Salcedo and N. Breton, Int. J. Mod. Phys. A \textbf{15}, 4341 (2000).

\bibitem{Camara} C. S. Camara, M. R. de Garcia Maia, J. C. Carvalho and J. A. S. Lima, Phys. Rev. D \textbf{69}, 123504 (2004).

\bibitem{Elizalde} E. Elizalde, J. E. Lidsey, S. Nojiri and S. D. Odintsov, Phys. Lett. B \textbf{574}, 1 (2003).

\bibitem{Novello4} M. Novello, E. Goulart, J. M. Salim and S. E. Perez Bergliaffa, Class.\ Quant.\ Grav.\  \textbf{24}, 3021 (2007).

\bibitem{Novello5} M. Novello, S. E. Perez Bergliaffa and J. M. Salim, Phys. Rev. D \textbf{69}, 127301 (2004).

\bibitem{Vollick} D. N. Vollick, Phys. Rev. D \textbf{78}, 063524 (2008).

\bibitem{Montiel} A. Montiel, N. Bretón and V. Salzano, Gen. Rel. Grav. \textbf{46}, 1758 (2014).

\bibitem{Breton} N. Breton, R. Lazkoz and A. Montiel, JCAP \textbf{10}, 013 (2012).

\bibitem{Bamba:2008ja}
  K.~Bamba and S.~D.~Odintsov,
  JCAP {\bf 0804}, 024 (2008).

\bibitem{Bamba:2008xa}
  K.~Bamba, S.~Nojiri and S.~D.~Odintsov,
  Phys.\ Rev.\ D {\bf 77}, 123532 (2008).

\bibitem{Bamba:2018cup}
  K.~Bamba, S. Nojiri and S.~D.~Odintsov,
  arXiv:1804.02275 [gr-qc], accepted for publication in Phys.\ Rev.\ D.

\bibitem{MRTB}
B.~Mishra, P.~P.~Ray, S.~K.~Tripathy and K.~Bamba,
submitted for publication.

\bibitem{Kruglov} S. I. Kruglov, Ann. Phys. \textbf{353}, 299 (2015).

\bibitem{Tolman} R. Tolman and P. Ehrenfest, Phys. Rev. \textbf{36}, 1791 (1930).

\bibitem{Planckm} M. Planck, The Theory of Radiation, Dover (1959) (translated from 1906).

\bibitem{Lemoine} D. Lemoine and M. Lemoine, Phys. Rev. D \textbf{52}, 1955 (1995).

\bibitem{Ellis} G. F. R. Ellis, R. Maartens and M. A. H. MacCallum, Gen. Rel. Grav. \textbf{39}, 1651 (2007).

\bibitem{Hawking} S. W. Hawking and G. F. R. Ellis, The large scale structure of space-time, (Cambridge, England, 1973).

\bibitem{Nojiri1} S. Nojiri, S. D. Odintsov and S. Tsujikawa, Phys. Rev. D \textbf{71}, 063004 (2005).

\bibitem{Barrow} J. D. Barrow, Class. Quant. Grav. \textbf{21}, L79 (2004).

\bibitem{Stefancic} H. Stefancic, Phys. Rev. D \textbf{71}, 084024 (2005).

\bibitem{Brevik} I. H. Brevik and O. Gorbunova, Gen. Rel. Grav, \textbf{37}, 2039 (2005).

\bibitem{Dabrowski} M. P. Dabrowski, Phys. Lett. B \textbf{659}, 184 (2005).

\bibitem{Bouhmadi} M. Bouhmadi-Lopez, P. F. Gonzalez-Diaz and P. Martin-Moruno, Phys. Lett. B \textbf{659}, 1 (2008).

\bibitem{Lazkoz} L. Fernandez-Jambrina, R. Lazkoz, J. Phys. Conf. Ser. \textbf{229}, 012037 (2010).











%
%
%
%
%
%
%






\end{thebibliography}
\end{document}